\documentclass[english,twocolumn,aps,prl,superscriptaddress,letter]{revtex4}
\usepackage[T1]{fontenc}
\usepackage[latin1]{inputenc}
\usepackage{graphicx}
\usepackage{amssymb}

\makeatletter

\makeatletter

\usepackage{times}

\newcommand{\op}[1]{\widehat{#1}}
\newcommand{\dagop}[1]{\widehat{#1}^{\dagger}}

\makeatother

\usepackage{babel}
\makeatother
\begin{document}

\title{Correlations in a BEC Collision: First-Principles Quantum Dynamics
with 150 000 Atoms}

\author{P. Deuar}

\email{pdeuar@science.uva.nl}

\affiliation{Van der Waals-Zeeman Instituut, Universiteit van Amsterdam, 1018
XE Amsterdam, Netherlands}

\affiliation{ARC Centre of Excellence for Quantum-Atom Optics, School of Physical
Sciences, The University of Queensland, Brisbane, QLD 4072, Australia}

\author{P. D. Drummond}

\affiliation{ARC Centre of Excellence for Quantum-Atom Optics, School of Physical
Sciences, The University of Queensland, Brisbane, QLD 4072, Australia}

\begin{abstract}
The quantum dynamics of colliding Bose-Einstein condensates with 150 000
atoms are simulated directly from the Hamiltonian using the stochastic
positive-\textsl{P} method. Two-body correlations between the scattered atoms
and their velocity distribution are found for experimentally accessible
parameters. Hanbury Brown--Twiss or thermal-like correlations are seen
for copropagating atoms, while number correlations for counterpropagating
atoms are even stronger than thermal correlations at short times.
The coherent phase grains grow in size as the collision progresses
with the onset of growth coinciding with the beginning of stimulated
scattering. The method is versatile and usable for a range of cold
atom systems. 
\end{abstract}


\maketitle
The prediction of many-body quantum dynamics is a long term goal of
investigation in a variety of scientific fields ranging from physics
to chemistry, biology and computation theory. It is a pivotal problem
for interacting systems, but challenging because of the complexity
of a full description of a quantum system, in which the number of
basis states grows exponentially with the number of particles. Experiments
with Bose-Einstein condensates of ultra-cold atoms give excellent
examples of phenomena that are not well described by standard approximations
such as the Gross-Pitaevskii (GP) equation. This equation treats the
macroscopically occupied wavefunction, but neglects atomic correlations
and fluctuations\cite{Dalfovo} which are especially prominent in
strongly interacting or dimensionally reduced gases, and in condensate
collisions. In the latter case, the GP equation fails because the
scattering initially occurs spontaneously into unoccupied modes, which
are ignored by a macroscopic wavefunction approach. Later, the scattering
becomes Bose-enhanced, and a coherent, non-perturbative treatment
of the scattered modes is essential.

Treatments of BEC collisions have included a slowly-varying envelope
approximation (SVEA) which estimates the scattering cross-section\cite{svea},
perturbation theory\cite{Zin,Bach}, and the semi-classical truncated
Wigner method\cite{Norrie12}. The last method is nonperturbative,
works well in one dimension\cite{DH93}, and appears to treat both
the initial spontaneous scattering and the later Bose enhancement.
However, we will show that it gives strongly incorrect results in
3D at large momentum cutoff, because the equations of motion are truncated.
Hence, there is a strong incentive to develop a quantitative, first-principles
method for these cases. 

This Letter also has a broader focus than just BEC. While path-integral
Monte Carlo methods are now very successful for calculating equilibrium
properties, quantum dynamics is not amenable to these techniques because
of the very rapid dephasing between different paths\cite{MonteCarlo}.
Phase-space distribution methods (such as the Glauber-Sudarshan\cite{GSp},
positive-\textsl{P}\cite{pp}, stochastic wavefunction\cite{Carusotto}, gauge-\textsl{P}\cite{UQbt})
do not suffer from this problem, and yet the scaling is still is only
linear in the system size. They have been applied successfully to
cold atom quantum dynamics in increasingly large systems, including
simulation of evaporative cooling to form a BEC\cite{Corney}, spin
squeezing and formation of two-component BECs\cite{Molmer}, correlation
dynamics in a uniform gas\cite{dynamix1}, the quantum evolution of
Avogadro's number of interacting atoms\cite{UQprl}, the dynamics
of atoms 
in a 1D trap\cite{Carusotto}, and molecular down-conversion\cite{Savage}.

Here we demonstrate the maturity and ready-to-use nature of
the original positive \textsl{P} method for truly macroscopic systems.
We simulate an average of 150 000 atoms, requiring $M=1.08\times10^{6}$
momentum modes. Since each of $M$ modes can have up to about $N$
atoms , the full Hilbert space contains at least $D\approx M^{N}\approx10^{1\,000\,000}$
orthogonal quantum states (or $D\approx10^{200\,000}$ if fixed total
atom number is assumed). This is one of the largest Hilbert spaces
ever treated in a first principles quantum dynamical simulation ---
made possible by probabilistic sampling rather than brute-force diagonalization.

The use of such a first principles, yet stochastic, simulation confers
several advantages in comparison with approximate methods\textit{.
Firstly}, all uncertainty in the results is confined to random statistical
fluctuations, with no systematic bias. This uncertainty can be reduced
by averaging over more stochastic realizations, and even more importantly,
can be reliably estimated from their spread. \textit{Secondly}, these
methods lead to relatively simple equations of motion, which are easily
adapted to realistic modeling of trap potentials and local losses.

We consider the collision of two pure $^{23}$Na BECs, with a similar
design to a recent experiment at MIT\cite{Vogels}. A $1.5\times10^{5}$
atom condensate is prepared in a cigar-shaped magnetic trap with frequencies
20 Hz axially and 80 Hz radially. A brief Bragg laser pulse coherently
imparts a velocity of ${2{\rm v_{Q}}}=19.64$ mm/s to half of the
atoms, much greater than the sound velocity of $3.1$ mm/s. At this
point the trap is turned off so that the wavepackets collide freely.
In a center-of-mass frame, atoms are scattered preferentially into
a spherical shell in momentum space with mean velocities ${\rm v_{s}}\approx{\rm v_{Q}}$.
As the density of atoms in this shell builds up, Bose-enhancement
of scattering into it is expected to begin. Of particular interest
are the distribution of scattered atom velocities, and correlations
between those atoms, which were recently shown to be experimentally
measurable\cite{JinBlochAspect}.

In present BEC experiments, the system can be described to a high
accuracy by the local interaction Hamiltonian\cite{Leggett}: \begin{equation}
\op{H}=\int\,\left[\frac{\hbar^{2}}{2m}\nabla\dagop{\Psi}\nabla\op{\Psi}+\frac{g}{2}\op\Psi^{\dagger2}\op\Psi^{2}\right]\, d\,^{3}\vec{x}\,.\label{H}\end{equation}
 The operator $\dagop{\Psi}(\vec{x})$ creates a bosonic atom at position
$\vec{x}=({\rm x,y,z})$ and obeys commutation relations $[\op{\Psi}(\vec{x}),\dagop{\Psi}(\vec{y})]=\widetilde{\delta}\,^{(3)}(\vec{x}-\vec{y})$,
with $\widetilde{\delta}$ a delta-function tempered by a momentum
cutoff $\vec{k}_{{\rm max}}$. The coupling constant $g$ depends
on the s-wave scattering length $a$ ($2.75$nm in the case of $^{23}$Na),
and for $\vec{k}_{{\rm max}}\ll1/a$ one finds that $g=4\pi\hbar^{2}a/m$.

 To calculate time-evolution, we employ the positive \textsl{P} representation\cite{pp,dynamix1}
because it preserves the full quantum dynamics. This approach utilizes
the completeness of the coherent-state basis\cite{GSp}. The density
matrix $\op{\rho}$ is expanded as a positive distribution $P$ over
off-diagonal coherent-state projectors\cite{pp}, thus preserving
quantum correlations: \mbox{$\op{\rho}=\int P(\vec{\alpha},\vec{\beta})\ d^{2M}\,\vec{\alpha}\, d^{2M}\vec{\beta}\ \left|\vec{\alpha}\right\rangle \langle\vec{\beta}^{*}|/[\langle\vec{\beta}^{*}\left|\vec{\alpha}\right\rangle ]$}.
Here $\op{\Psi}(\vec{x})=\sqrt{1/V}\sum_{\vec{k}}e^{-i\vec{k}\cdot\vec{x}}a_{\vec{k}}\,$
for momentum-mode operators $a_{\vec{k}}$ in a volume $V$. The coherent
state $\left|\vec{\alpha}\right\rangle =\otimes_{\vec{k}}\left|\alpha_{\vec{k}}\right\rangle $
is a joint eigenstate of each $a_{\vec{k}}$, with complex eigenvalue
$\alpha_{\vec{k}}$\cite{GSp}. When used to expand the master equation
$i\hbar\partial\op{\rho}/\partial t=[\op{H},\op{\rho}\,]$\,\cite{dynamix1},
this leads to a Fokker-Planck or diffusion equation in the probability
$P$, which  is equivalent to solving an ensemble of stochastic equations
for the sampled variables $\vec{\alpha}$ and $\vec{\beta}$. The
equations are simplified on discrete Fourier transforming to a conjugate
spatial lattice $\vec{x}$, where $\alpha_{\vec{x}}=\sum_{\vec{k}}\alpha_{\vec{k}}\exp(i\vec{k}\cdot\vec{x})/\sqrt{M}$:
\begin{eqnarray}
i\hbar\frac{d\alpha_{\vec{x}}}{dt} & = & \left[-\frac{\hbar^{2}}{2m}\nabla^{2}+\frac{g}{\Delta V}\,\alpha_{\vec{x}}\beta_{\vec{x}}+\sqrt{i\hbar g}\ \xi_{\vec{x}}\right]\alpha_{\vec{x}}\label{eqnmotion}\\
-i\hbar\frac{d\beta_{\vec{x}}}{dt} & = & \left[-\frac{\hbar^{2}}{2m}\nabla^{2}+\frac{g}{\Delta V}\,\alpha_{\vec{x}}\beta_{\vec{x}}+\sqrt{-i\hbar g}\ \widetilde{\xi}_{\vec{x}}\right]\beta_{\vec{x}}.\nonumber \end{eqnarray}
 Here $\nabla^{2}\alpha_{\vec{x}}$ is the discretized analogue of
$\nabla^{2}\alpha(\vec{x})$ for a field, and $\xi_{\vec{x}}$ and
$\widetilde{\xi}_{\vec{x}}$ are real Gaussian noises, independent
at each time step (of length $\Delta t$) and lattice point, with
standard deviations $1/\sqrt{\Delta V\Delta t}$, where $\Delta V=V/M$
. 

There is an equivalence between statistical averages of moments of
$\alpha_{\vec{k}}$ and $\beta_{\vec{k}}$, and corresponding normally-ordered
expectation values of operators $a_{\vec{k}},{\,\, a}_{\vec{k}}^{\dagger}$.
As the number of trajectories, $S$, grows towards $\infty$, the
correspondence becomes exact. These stochastic equations are just
the mean-field GP equations in a doubled phase-space, plus noise terms.
Remarkably, these modifications incorporate all effects beyond the
GP equation, provided certain phase-space boundary conditions are met\cite{UQbt,dynamix2}.

Uncertainty in the observables is estimated by binning trajectories,
then calculating the observable predictions from each bin, and using
the central limit theorem to estimate the standard deviation in the
final mean of bin means. Lattice spacings $\Delta t$ and $\Delta\vec{x}$
are chosen by reducing them until no further change is seen. In the
figures, results are presented in terms of velocity space, $\vec{v}=\hbar\vec{k}/m$,
the Fourier transformed field $\op{\Psi}(\vec{v}),$ and the velocity
space density $\rho(\vec{v})=\langle\dagop{\Psi}(\vec{v})\op{\Psi}(\vec{v})\rangle$.

Following earlier procedures\cite{DC}, we discretize onto a
$M=432\times50\times50$ lattice with $k_{{\rm x,max}}=1.4\times10^{7}$/m
and $k_{{\rm y,z,max}}=6.2\times10^{6}$/m. We begin the simulation
in the center-of-mass frame at the moment the lasers and trap are
turned off \mbox{($t=0$).} The $k_{{\rm max}}$ and lattice size
are chosen large enough to encompass all relevant phenomena but small
enough that the spacing ($\pi/k_{{\rm max}}$) is much larger than
$a$. The initial wavefunction is modelled as the GP solution of the
trapped $t<0$ condensate, but modulated with a factor $\left[e^{ik_{{\rm Q}}{\rm x}}+e^{-ik_{{\rm Q}}{\rm x}}\right]/\sqrt{2}$
which imparts initial velocities ${\rm v_{x}}={\rm \pm v_{Q}}=\pm\hbar k_{{\rm Q}}/m$
in the ${\rm x}$ direction. For computational reasons, the mean number
of atoms in the system is $1.5\times10^{5}$ here, compared to $\approx3\times10^{7}$
in the MIT experiment\cite{Vogels}. As in other recent treatments\cite{Norrie12},
we ignore thermal atoms and initial quantum depletion, for simplicity.
For our parameters, 10\% thermal component will occur at $\approx0.38T_{c}$\cite{Dalfovo},
giving a $\approx1\%$ quantum depletion of the ground state in the
center of the cloud\cite{Depletion}. These small corrections can
be included in the initial state\cite{Ruostekoski}.

\begin{figure}
\begin{centering}\includegraphics[width=5.25cm]{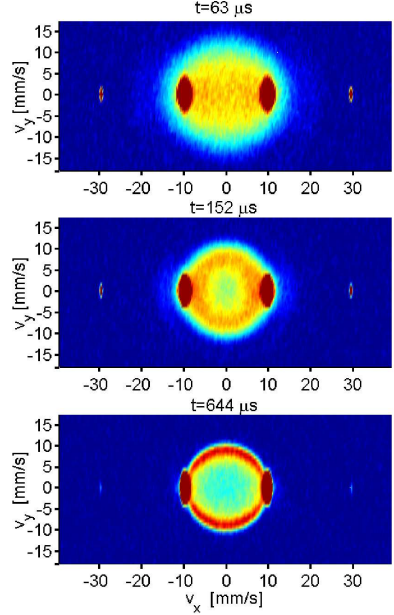}\hspace*{0.8cm} \par\end{centering}

\vspace*{-10pt}

\caption{\label{figa} \textbf{Evolution of velocity distribution.} The distribution
$\rho({\rm {v_{x}},{\rm {v_{y}})}}$ has been integrated over one
transverse dimension ${\rm z}$. Color from blue to red indicates
increasing density (its range varies between panels). $S=2048$ trajectories. }

\vspace*{-10pt}

\end{figure}

%
\begin{figure}
\begin{centering}\includegraphics[width=200pt]{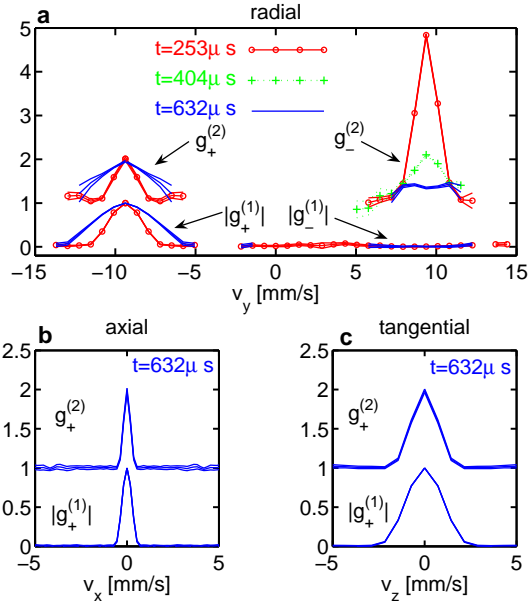} \par\end{centering}

\vspace*{-10pt}

\caption{\label{figb} \textbf{Correlations between scattered atoms.} All
are between scattered atoms at a maximum density point in the shell
(with velocity \mbox{$\vec{v}_{0}=(0,-9.37,0)$ mm/s} relative to
the COM) and those with a shifted velocity \mbox{$\vec{v}$}, where:
\textbf{a}: $\vec{v}=(0,{\rm v_{y}},0)$, \textbf{b}: $\vec{v}=({\rm v_{x}},0,0)$,
and \textbf{c}: $\vec{v}={\rm (0,0,{\rm {v_{z}})}}$. To reduce statistical
noise, averages $g_{\pm}^{(n)}=\frac{1}{V_{0}}\int g^{(n)}(\vec{v}_{0}+\vec{\delta v},\vec{v}\pm\vec{\delta v})\, d^{2}\vec{\delta v}$,
over a volume $V_{0}$ in velocity space are plotted\cite{vol}. Triple
lines are $1\sigma$ errors.}

\vspace*{-10pt}

\end{figure}

%
\begin{figure}
\begin{centering}\includegraphics[clip,width=8cm]{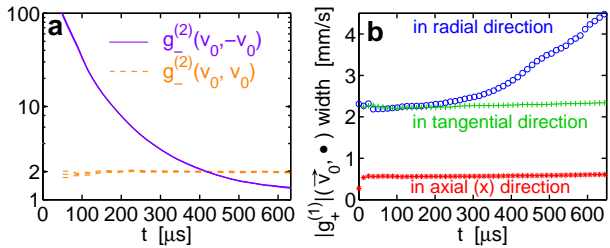} \par\end{centering}

\vspace*{-10pt}

\caption{\label{figc} \textbf{Evolution of correlations between scattered
atoms.} Parameters and $\vec{v}_{0}$ as in Fig.~\ref{figb}. \textbf{b}:
The Full-width half-maximum (FWHM) width of $|g_{\pm}^{(1)}|$ in
velocity space. }

\vspace*{-10pt}

\end{figure}

Figure~\ref{figa} shows the formation of the scattered atom shell.
Careful inspection shows that the mean scattered atom speed ${\rm v_{s}}$
is less than the wavepacket speed ${\rm |v_{Q}|}$, as noted in \cite{Bach}.
We also see weak scattering between two atoms from a wavepacket at
$\pm{\rm v_{Q}}$ to one atom at $\pm3{\rm {v_{Q}}}$ and one at $\mp{\rm {v_{Q}}}$.

Ranged two-body correlations give insight into typical small-scale
behaviour during a single experimental run. The first-order correlation
function $g^{(1)}(\vec{v}_{1},\vec{v}_{2})=\langle\dagop{\Psi}(\vec{v}_{1})\op{\Psi}(\vec{v}_{2})\rangle/\sqrt{\rho(\vec{v}_{1})\,\rho(\vec{v}_{2})}$,
describes coherence between particles with velocity $\vec{v}_{1}$
and $\vec{v}_{2}$. The second-order (number) correlation function
$g^{(2)}(\vec{v}_{1},\vec{v}_{2})=\langle\dagop{\Psi}(\vec{v}_{1})\dagop{\Psi}(\vec{v}_{2})\op{\Psi}(\vec{v}_{1})\op{\Psi}(\vec{v}_{2})\rangle/\rho(\vec{v}_{1})\,\rho(\vec{v}_{2})$
gives the average shape and size of {}``lumps'' in the velocity
distribution. 

\begin{figure}
\begin{centering}\includegraphics[width=8.5cm]{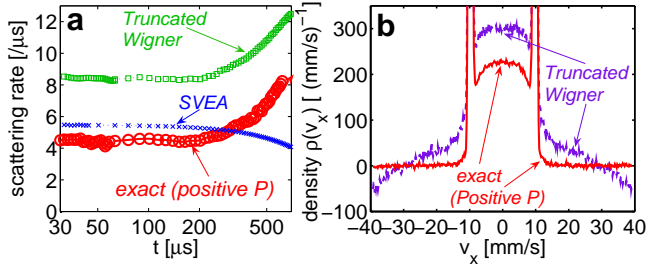}\par\end{centering}

\mbox{}\vspace*{-10pt}

\caption{\label{figwig} \textbf{Comparison of exact and approximate methods}.
Panel \textbf{a}: \textit{Rate of scattering out of coherent wavepackets.}
Obtained by counting atoms outside of spheroidal regions covering
the coherent wavepackets centered at $\pm{\rm v_{Q}}$ and $\pm3{\rm v_{Q}}$.
Panel \textbf{b}: \textit{Distribution of axial (}x\textit{) velocity
of scattered atoms} at $t=657\mu$s; In both panels: $S=2048$ (Pos.
P), $S=672$ (Wigner). }

\vspace*{-15pt}

\end{figure}

The dynamics of the correlations among scattered atoms are shown in
Figs.~\ref{figb} and~\ref{figc}. Locally the atoms are thermally
bunched with $g^{(2)}(\vec{v},\vec{v})\approx2$ in a {}``Hanbury Brown--Twiss''
manner (Fig.~\ref{figb}). This behaviour has been confirmed qualitatively
in a similar recent He$^{*}$ experiment\cite{Westbrook}. The local
region over which coherence is strong, dubbed a {}``phase grain''
by Norrie \textit{et al}\cite{Norrie12}, is described by $|g^{(1)}|$.
It closely matches the condensate wavepackets' $\rho(\vec{v})$ in
size, and is wider than $g^{(2)}$ by $\approx\sqrt{2}$ (Fig.~\ref{figb}).
We find that the orientation of these phase grains is constant throughout
the whole spherical shell. Interestingly, after $t\approx200\,\mu$s,
the phase grains expand significantly in the radial direction (relative
to COM) (Figs.~\ref{figb}\textbf{a} and~\ref{figc}\textbf{b}).
This onset of growth coincides with the beginning of Bose stimulated
scattering (see below and Fig.~\ref{figwig}\textbf{a}, circles).

Atoms with velocities $\vec{v}$ and $-\vec{v}$ on opposite sides
of the spherical shell are not coherent ($|g^{(1)}|\approx0$), but
are correlated in number (Fig.~\ref{figb}\textbf{a}). Initially,
correlations are extreme: $g^{(2)}(k,-k)\gg2$. This is analogous
to a two-mode mixed state $p\,|1,1\rangle\langle1,1|+(1-p)\,|0,0\rangle\langle0,0|$
with a small probability $p\,$ of single atoms in both modes and
otherwise vacuum. There $g^{(2)}=1/p$. At longer times, $g^{(2)}$
is seen to decay in Fig.~\ref{figc}\textbf{a}, although it is still
much greater than the thermal value of two for $t\gtrsim200\mu$s
when the phase grain contains several atoms. To measure short time
velocity correlations, one might try to preserve them by suddenly
switching off the atomic interactions using a Feshbach resonance during
the collision. After expansion, they would develop into position correlations\cite{JinBlochAspect}.

Some previous correlation estimates are in qualitative agreement:
For longer times, $g^{(2)}(\vec{v},\vec{v})=2$, as well as $g^{(2)}(\vec{v},-\vec{v})\approx2$
and $g^{(1)}(\vec{v},-\vec{v})\approx0$ were predicted\cite{Zin}.
Truncated Wigner calculations\cite{Norrie12} saw the presence of
phase grains, but their orientation or dynamics were not studied.
High initial correlations may have not been seen due to the known
poor signal-to-noise ratio in that method.


The scattering rate (Fig.~\ref{figwig}\textbf{a}, circles) goes
through two distinct phases: The \textit{spontaneous regime} of constant
scattering into almost empty modes is seen for \mbox{$30\,\mu$s$\,\lesssim t\lesssim200\,\mu$s},
followed by the \textit{stimulated (Bose-enhanced)} regime for times
\mbox{$t\gtrsim200\,\mu$s}, where there is a decided increase in
scattering rate despite a lessening overlap between the colliding
wavepackets. We interpret this transition as the onset of Bose enhancement
of scattering into the spherical shell around $|\vec{v}|\approx{\rm v_{s}}=9.37$
mm/s. As a rough check, it should begin when the number of particles
in a locally coherent region ({}``phase grain'') approaches one.
Using the widths of $g^{(1)}$ from Fig.~\ref{figc}\textbf{b}, and
the calculated density at $|\vec{v}|={\rm v_{s}}$ one finds $\approx0.9$
atoms per phase grain at $200\,\mu$s.

%

A comparison to approximate methods used previously is instructive.
Fig.~\ref{figwig} shows our total predicted scattering rate and
the distribution of axial (x) velocities, compared to the approximate
truncated Wigner method. The accuracy of that method is very poor
with our parameters, \textit{even surprisingly so}. It adds a halo
of false particles (detail in Fig~\ref{figwig}\textbf{b}) out to
about $\pm2{\rm v_{Q}}$, while at higher velocities unphysical negative
densities are obtained. Since it is a hidden-variable theory it must
introduce half a virtual particle per mode in the initial conditions
to model vacuum fluctuations, but it does not distinguish them from
the {}``real particles''. Then, virtual particles at velocity $\vec{v}$
are scattered by the condensates at $\approx\pm\vec{v}_{{\rm Q}}$
in the process $\vec{v}\ \&\ \pm\vec{v}_{{\rm Q}}\ \to\ \vec{v}\,'\ \&\ (\vec{v}\pm\vec{v}_{{\rm Q}}-\vec{v}\,')$.
As a result, modes at high velocities become depleted compared to
the physical vacuum, while the extracted virtual particles accumulate
at lower velocities and take on the appearance of a real density,
as was also discussed previously\cite{Sinatra}.

For any single momentum mode this effect is small, but it becomes
very significant when a large number of modes are calculated. A relatively
higher momentum cutoff will increase the error, as the fraction of
virtual particles increases (or vice-versa\cite{Norrie12}). This
indicates a generic ultra-violet divergence of the error with the
truncated Wigner method. 

The main limitation of the positive-\textsl{P} method is the growth of sampling
uncertainty with time. It eventually reaches a size where it is no
longer practical to produce enough trajectories for useful precision.
In our case useful results are obtained for $t\lesssim660\,\mu$s.
This useful time range depends on several factors, with coarser lattices,
weaker interactions, or smaller density all extending it\cite{dynamix1}.
Significant extensions appear achievable by tailoring appropriate
stochastic gauges\cite{dynamix2,UQbt} or basis sets to particular
systems. 

In conclusion, we have simulated the quantum dynamics of macroscopic
interacting Bose gases from first-principles, obtaining momentum space
densities and ranged correlation functions for atoms scattered during
the collision of two BECs. Previous approximate calculations were
partly verified, while a variety of new phenomena are also predicted,
including the growth of phase grains in the radial momentum direction,
and strong correlations at short times between scattered atom pairs.
The truncated Wigner method was confirmed strongly incorrect in regimes
where the number of condensed atoms per lattice site is less than
one.

This demonstrates that phase-space methods are a tool that is ready-to-use
for first principles calculations for experimentally realizable systems.
Similar calculations appear feasible for a broad range of cold atom
systems (including fermions\cite{CorneyDrummond}). A range of other
phenomena that are difficult to describe quantitatively with approximate
methods (e.g. macroscopic EPR and entanglement\cite{Macro}) may be
accessible with this approach.

\begin{acknowledgments}
We thank G.~Shlyapnikov, J.~Chwede\'{n}czuk, M.~Trippenbach, P.~Zi\'{n},
K. Kheruntsyan and C.~W.~Gardiner for valuable discussions. The
work was supported financially by the Australian Research Council,
as well as by the NWO as part of the FOM quantum gases project. 
\end{acknowledgments}


\begin{thebibliography}{10}
\bibitem{vol} These are with $\vec{\delta v}$ lying in the volume
$V_{0}$, which had dimensions $3.82\times5.05$\,mm/s in the x and
z directions, respectively, and was chosen small enough that $g^{(n)}$
does not vary significantly within it, as evidenced by $\partial g_{\pm}^{(n)}/\partial V_{0}\approx0$.
\end{thebibliography}
\end{document}